\documentclass[12pt,twoside,a4paper,fleqn]{article}
\usepackage[left=25mm,top=20mm,right=14mm,bottom=30mm]{geometry}

\usepackage{epsfig}
\usepackage{graphicx}
\usepackage{amssymb}
\usepackage{mathrsfs}
\usepackage{multirow}
\usepackage{multicol}

\newcommand{\pr}{\partial}

\newcommand{\rta}{\rightarrow}

\newcommand{\upa}{\uparrow}
\newcommand{\dwa}{\downarrow}
\newcommand{\ep}{\epsilon}

\newcommand{\p}{\prime}
\newcommand{\om}{\omega}

\newcommand{\beq}{\begin{equation}}
\newcommand{\eeq}{\end{equation}}

\newcommand{\ball}{\begin{align}}
\newcommand{\eall}{\end{align}}

\newcommand{\beqar}{\begin{eqnarray}}
\newcommand{\eeqar}{\end{eqnarray}}

\newcommand{\ben}{\begin{enumerate}}
\newcommand{\een}{\end{enumerate}}
\newcommand{\mud}{\mu_{d}}
\newcommand{\mus}{\mu_{s}}
\makeatletter
\newcommand*{\rom}[1]{\expandafter\@slowromancap\romannumeral #1@}
\makeatother

\begin{document}
\title{ Electrical resistivity in 2d Kondo lattice systems}
\author{ Komal Kumari\\ \small Department~of~Physics,~Himachal~Pradesh~University,\\ \small~Shimla,~India, Pin:171005.\\ \small komal.phyhpu@gmail.com; sharmakomal611@gmail.com}


\date{}
\maketitle

\begin{abstract}
I extend the calculations represented in \cite{konav} regarding the resistivity in Kondo lattice materials from $3d$ syatem to $2d$ systems. In the  present work I consider a 2d system, and  memory function is computed. However, results found in 2d  case are different from 3d system . I find that in $2d$ in low temperature regime($ k_{B}T\ll \mud$) resistivity shows power law($\frac{1}{T}$) behaviour and in the high temeprature regime($ k_{B}T\gg\mud$) resistivity varies linearly with temperature. In $3d$ these behaviours are as $\frac{1}{T}$ and as $T^{\frac{3}{2}}$ respectively.

\end{abstract}

\section{Introduction}
The electrical resistivity originates from spin dependent scattering processes  \cite{andersonbook,vath,jk,konav,mathon} . These
 scattering events are considered between  the localised $d$ or $f$-moments and conduction $s$-electron. The arrangements of spins of localized moments and mobile $s$-electrons  construct Kondo-lattice system. In Kondo system at each lattice site a local moment interact via exchange coupling with the spin of conduction electron\cite{jk,konav,mathon,scholarpedia,kondo,hewson,kumh,shenoy}.   These conduction electrons spins undergo the spin-flip scattering processes with the localized magnetic moment spins\cite{kasuya,mannari,weiss}, thus this mechanism leads to resistivity.\\
 
The present work is dedicated to the  compution of  the  electrical resistivity of $2d$ Kondo lattice system.  I have found analytically that resistivity $\rho \propto \frac{1}{T}$ in low temperature regime ($ k_{B}T\ll \mud$). However, in  the high temperature regime($ k_{B}T\gg \mud$) $\rho \propto T$.\\

  The recent published paper\cite{konav} has reported  the resistivity  in  the $3d$ Kondo lattice materials using the memory function formalism.
The electrical  resistivity in  such system shows power law behaviour ($\rho \propto \frac{1}{T}$) in the low temperature  ( $ k_{B}T\ll \mud$)  limit. In the high temperature regime ($ k_{B}T\gg \mud$) resistivity scales to cube root of T($\rho\propto T^{\frac{3}{2}}$). The coupling Hamiltonian in the Kondo lattice system  is the $s$-$d$ Hamiltonian\cite{mathon}:



\beqar
H_{sd}&=&\frac{J}{N}\sum_{k^{\p}k} \bigg\{ a^{\dagger}_{k^{\p}\uparrow} a_{k\downarrow}S^{-}(k^{\p}-k)+ a^{\dagger}_{k^{\p}\downarrow} a_{k\uparrow}S^{+}(k^{\p}-k)+(a^{\dagger}_{k^{\p}\uparrow}
a_{k\upa}-a^{\dagger}_{k^{\p}\downarrow} a_{k\dwa})S^{z}(k^{\p}-k)\bigg\}\nonumber\\&& \label{sd1}
\eeqar
Here $a^{\dagger}_{k^{\p}\uparrow}( a_{k\downarrow}) $ are creation(annihilation) operators of $s$-electrons and $S^{-}(k^{\p}-k)$ is spin lowering operator of quasi $d$ or $f$-electrons. 
In the next section, I write directly imaginary part of  the memory function(all the mathematical details are available in ref.\cite{konav}). 

\section{General expression of Memory function Formula}

Writing the  imaginary part of the  memory function formula\cite{konav,singh,gotze,luxmi,bhalla} for $s$ and $d$ or $f$ electrons

\beqar
M^{\p\p}(\om)&=&\frac{J^2 m\pi}{N^2\hbar^3 n  V \om} \sum_{k^{\p}k}(v_{1}(k^\p)-v_{1}(k))^2\{  f^s_{k^\p}(1-f^s_{k})\sum_{k_{d},k^{\p}_{d}} (f^{d}_{k_{d}}-f^{d}_{k^{\p}_{d}})-\nonumber\\&&~~~~~~~~~~~~(f^s_{k}-f^s_{k^\p}) \sum_{k_{d},k^{\p}_{d}}f^{d}_{k_{d}}(1-f^d_{k^{\p}_{d}})\} \nonumber\\&&~~~~~~~~~~~~~~~~~~[\delta(\frac{\ep_{k^\p}}{\hbar}-\frac{\ep_{k}}{\hbar}-\om_{k^\p-k}+\om)-\delta(\frac{\ep_{k^\p}}{\hbar}-\frac{\ep_{k}}{\hbar}-\om_{k^\p-k}-\om)]. \label{m2} \nonumber\\&&
\eeqar 
Replacing velocity components for states $ k$ and $k^\p$ in terms of wavevectors ($v_{1}(k^\p)=\frac{\hbar}{m}k^\p)$. Next,  write the momentum conservation $\vec{k}^\p-\vec{k}=\vec{k}^{\p}_{d}-\vec{k}_{d}=\vec{q}$.  Insert  an integral $dq \delta (\vec{q}-|\vec{k}^\p-\vec{k}|)$ over $q$ into equation (\ref{m2}), which simplifies the magnitude of $(\vec{k}^\p-\vec{k})$ greatly. The spatial isotropy of the free electron writes the velocity as $\frac{v^2}{3}=(v^2_{x}+v^2_{y}+v^2_{z})$.  Finally, converting the  summations into integrals for $k$ and $k^\p$ using $\frac{1}{A}\sum\rta\int\frac{ d^2k}{(2\pi)^2}$, one gets
\beqar
M^{\p\p}(\om)&=&\frac{J^2\pi V }{3N^2 m n  }\int_{0}^{\infty}\frac{ dq}{\om} ~q^2\int_{0}^{\infty}  \frac{d^2k}{(2\pi)^2}\int_{0}^{\infty}\frac{d^2k^\p}{(2\pi)^2} \delta (\vec{q}-|\vec{k}^\p-\vec{k}|)\nonumber\\&&F(f^s_{k},f^s_{k^\p},f^{d}_{k_{d}},f^d_{k^{\p}_{d}})[\delta(\ep_{k+q}-\ep_{k}-\hbar\om_{q}+\hbar\om)-\delta(\ep_{k+q}-\ep_{k}-\hbar\om_{q}-\hbar\om)].\nonumber\\&& \label{m3}
\eeqar
The function $F(f^s_{k},f^s_{k^\p},f^{d}_{k_{d}},f^d_{k^{\p}_{d}})$ denotes the short hand notation for Fermi distribution functions. Writing the integral $\int d^2k=2\pi \int k dk$, $\int d^2k^\p= \int k^\p dk^\p\int_{0}^{2\pi}d\phi$ (take  $k$ as pointing along the $z-$direction) and expression (\ref{m3}) changes to
\beqar
M^{\p\p}(\om)&=&\frac{J^2\pi V }{3N^2 m n  }\frac{(2\pi)}{(2\pi)^4}\int_{0}^{\infty} \frac{ dq}{\om} ~q^2\int_{0}^{\infty} k dk \int_{0}^{\infty}k^{\p} dk^\p\nonumber\\&&~~~~~~~~~~\int_{0}^{2\pi} d\phi~ \delta (q-\sqrt{(k^{\p 2}+k^2-2k^{\p} k\cos\phi)})
\sum_{k_{d},k^{\p}_{d}}F(f^s_{k},f^s_{k^\p},f^{d}_{k_{d}},f^d_{k^{\p}_{d}})\nonumber\\&&~~~~~~~~~~~~~~[\delta(\ep_{k+q}-\ep_{k}-\hbar\om_{q}+\hbar\om)-\delta(\ep_{k+q}-\ep_{k}-\hbar\om_{q}-\hbar\om)]. \label{m4}
\eeqar 
The computation of $\phi$ integral(
appendix A) simplifies the eqn (\ref{m4}) as
\beqar
M^{\p\p}(\om)&=&\frac{J^2\pi V }{3N^2 m n  }\frac{(2\pi)}{(2\pi)^4}\int_{0}^{\infty} \frac{ dq}{\om} ~q^2 \int^{\infty}_{q_{0}}
\frac{k~dk}{\sqrt{k^2-q^2_{0}}} \int^{\infty}_{0} k^\p dk^\p F(f^s_{k},f^s_{k^\p},f^{d}_{k_{d}},f^d_{k^{\p}_{d}})\nonumber\\&&~~~~~~~~~~~~[\delta(\ep_{k+q}-\ep_{k}-\hbar\om_{q}+\hbar\om)-\delta(\ep_{k+q}-\ep_{k}-\hbar\om_{q}-\hbar\om)]. \label{m5}
\eeqar 

Converting $k$ and $k^\p$ integral into $\ep$ and $\ep^\p$ and replacing $\ep^\p$ integral from appendix B

\beqar
M^{\p\p}(\om)&=& \frac{J^2 2\pi A^2}{6N^2 n V (2\pi)^3}(\frac{m}{\hbar^4})(\frac{\hbar^2}{2m})^{\frac{1}{2}}\int_{0}^{q_{D}} dq q^2 \int_{\ep_{0}}^{\infty}\frac{d\ep}{\sqrt{\ep-\ep_{0}}}\frac{1}{\om}\bigg[\bigg\{f^{s} (\ep_{k}+\hbar\om_{q}-\hbar\om)-\nonumber\\&&~~~~~~~~~~~~~~~~~~f^{s} (\ep_{k}+\hbar\om_{q}+\hbar\om)\bigg\} \bigg(1-f^s(\ep_{k})\bigg)F^{1}_{d}(q)+\nonumber\\&&~~~~~~~~~~~~~~~~~~\bigg\{f^{s} (\ep_{k}+\hbar\om_{q}-\hbar\om)-f^{s} (\ep_{k}+\hbar\om_{q}+\hbar\om) \bigg\}F^{2}_{d}(q)\bigg].\nonumber\\&& \label{m6}
\eeqar
This is required general expression of imaginary part of the Memory Function, which is valid for all frequencies and all temperature regimes. Operating the limit $\om\rta0$, we rewrite equation (\ref{m6}) as follows:

\beqar
M^{\p\p}(\om=0,T)&=& \frac{J^2 2\pi A^2m2\beta\hbar}{6N^2 n V (2\pi)^3\hbar^3\sqrt{2m}}\int_{0}^{q_{D}} dq q^2 \int_{\ep_{0}}^{\infty}\frac{d\ep}{\sqrt{\ep-\ep_{0}}}\bigg\{f^s(\ep+\hbar\om_{q})(1-f^s(\ep+\hbar\om_{q})) \bigg\}\nonumber\\&&~~~~~~~~~~~~~~~~~~~\bigg[ (1-f^s(\ep))\sum_{k_{d},k^{\p}_{d}} (f^{d}_{k_{d}}-f^{d}_{k^{\p}_{d}})+\sum_{k_{d},k^{\p}_{d}}f^{d}_{k_{d}}(1-f^d_{k^{\p}_{d}})\}\bigg]. \nonumber\\&& \label{m7}
\eeqar
The use of valid reasonable assumptions simplify the above expression: (1) The $s$-electrons fermi functions  energy $k_{B}T\ll\mus$ (chemical potential for s-electrons,  $\mus\simeq 10 eV$), which is much greater than room temperature($\sim0.025eV$) (2)  the energy scale of magnetic excitation $\hbar\om_{q}\ll\mus$ ($meV$) is much less than the assumed  $s$ electrons chemical potential\cite{konav,mannari}. Using the second assumption  the Fermi function $f^s(\ep+\hbar\om_{q})=\frac{1}{e^{\beta(\ep+\hbar\om_{q}-\mu_{s})}+1}$ approximates  to $f^s(\ep)$ and the above expression becomes
\beqar
M^{\p\p}(\om=0,T)&=&\frac{p_{0}}{k_{B}T}\int_{\ep_{0}}^{\infty}\frac{d\ep}{\sqrt{\ep-\ep_{0}}} \frac{1}{\beta}\delta(\ep-\mus)\bigg[ \int_{0}^{q_{D}} dq q^2\sum_{k_{d},k^{\p}_{d}}\bigg\{ (1-f^s(\ep))(f^{d}_{k_{d}}-f^{d}_{k^{\p}_{d}}) \nonumber\\&&~~~~~~~~~~+ f^{d}_{k_{d}}(1-f^d_{k^{\p}_{d}})\bigg\}\bigg].  \label{m8}
\eeqar
For $\mus\gg\ep_{0}$ the above expression simplifies to  
\beqar
M^{\p\p}(\om=0,T)&=& \frac{p_{0}}{\sqrt{\mus}}\bigg[\frac{1}{2}\int_{0}^{q_{D}} dq q^2\sum_{k_{d},k^{\p}_{d}}(f^{d}_{k_{d}}-f^{d}_{k^{\p}_{d}})+\int_{0}^{q_{D}} dq q^2\sum_{k_{d},k^{\p}_{d}}f^{d}_{k_{d}}(1-f^{d}_{k^{\p}_{d}}) \bigg],  \label{m9}
\eeqar
here $p_{0}=\frac{J^2A^2m}{3N^2nV(2\pi)^2\hbar^2\sqrt{2m}}$. We take integral terms as $\mathbb{I}_{1}(q)$ and $\mathbb{I}_{2}(q)$ and substitute computed expressions from appendices $C$ and $D$(eqn \ref{iq1},\ref{iq22}) 

\beqar
M^{\p\p}(\om=0,T)&=& \frac{J^2 A^2 m}{12\pi^2N^2nV\hbar^3 q_{s}}\bigg[\frac{Aq^5_{D}}{40\pi}\bigg\{\beta\int_{0}^{\infty}\frac{d\ep_{d}e^{\beta(\ep_{d}-\mud)}}{(e^{\beta(\ep_{d}-\mud)}+1)^2}+\beta^2\int_{0}^{\infty}\frac{d\ep_{d}\ep_{d}e^{\beta(\ep_{d}-\mud)}}{(e^{\beta(\ep_{d}-\mud)}+1)^2}\nonumber\\&&~~~~~-2\beta^2\int_{0}^{\infty}\frac{d\ep_{d}\ep_{d}e^{2\beta(\ep_{d}-\mud)}}{(e^{\beta(\ep_{d}-\mud)}+1)^3}\bigg\}+\frac{Am\lambda q^3_{D}}{6\pi\hbar^2}\int_{0}^{\infty}\frac{d\ep_{d}e^{\beta(\ep_{d}-\mud)}}{(e^{\beta(\ep_{d}-\mud)}+1)^2}+\nonumber\\&&~~~~~~~~~~~~~~ \frac{Aq^5_{D}}{20\pi}\bigg\{\beta\int_{0}^{\infty}\frac{d\ep_{d}e^{\beta(\ep_{d}-\mud)}}{(e^{\beta(\ep_{d}-\mud)}+1)^3}+\beta^2\int_{0}^{\infty}\frac{d\ep_{d}\ep_{d}e^{\beta(\ep_{d}-\mud)}}{(e^{\beta(\ep_{d}-\mud)}+1)^3}\nonumber\\&&~~~~~~~~~~~~~~~~~~~~~~-2\beta^2\int_{0}^{\infty}\frac{d\ep_{d}\ep_{d}e^{2\beta(\ep_{d}-\mud)}}{(e^{\beta(\ep_{d}-\mud)}+1)^4}\bigg\}\bigg].  \label{m10}
\eeqar
In the above expression we have replaced $\sqrt{2m\mus}= \hbar q_{s}$ and $m_{d}=m\lambda$.
The transformation of the variables in all integrands $x=\beta(\ep-\mud)$ changes the above expression as

\beqar
M^{\p\p}(\om=0,T)&=& \frac{J^2 A^2 m}{12\pi^2N^2nV\hbar^3} \bigg[\frac{Aq^4_{s}}{40\pi}(\frac{q_{D}}{q_{s}})^5\bigg\{\int_{-\beta\mud}^{\infty}\frac{dx~e^{x}}{(e^{x}+1)^2}+\beta\int_{-\beta\mud}^{\infty}\frac{ dx(\frac{x}{\beta}+\mud)e^x}{(e^{x}+1)^2}\nonumber\\&&~~~-2\beta\int_{-\beta\mud}^{\infty}\frac{ dx(\frac{x}{\beta}+\mud)e^{2x}}{(e^{x}+1)^3}\bigg\}+\frac{Am\lambda}{6\pi\hbar^2}(\frac{q_{D}}{q_{s}})^3\frac{q^2_{s}}{\beta}\int_{-\beta\mud}^{\infty}\frac{dx~e^{x}}{(e^{x}+1)^2}+\nonumber\\&&~~~~~~~\frac{Aq^4_{s}}{20\pi}(\frac{q_{D}}{q_{s}})^5\bigg\{\int_{-\beta\mud}^{\infty}\frac{dx~e^{x}}{(e^{x}+1)^3}+\beta\int_{-\beta\mud}^{\infty}\frac{ dx(\frac{x}{\beta}+\mud)e^x}{(e^{x}+1)^3}\nonumber\\&&~~~~~~~~~~~~~~~~~~~~~-2\beta\int_{-\beta\mud}^{\infty}\frac{ dx(\frac{x}{\beta}+\mud)e^{2x}}{(e^{x}+1)^4}\bigg\}\bigg].  \label{m11}
\eeqar

Thus, we have obtained more simplified form of the expression computed under two important assumptions(mentioned as 1 and 2). We further analyse the general expression for low and high temperature limit.
\section{Special cases:}
\subsection{Low temperature limit $(k_{B}T\ll\mud)$}
For limit $\beta\mud\gg1$ the expression (\ref{m11}) reduces to
\beqar
M^{\p\p}(T\rta 0)&\simeq& \frac{J^2 A^2 m}{12\pi^2N^2nV\hbar^4} \bigg[ \frac{Aq^4_{s}}{40\pi}(\frac{q_{D}}{q_{s}})^5\bigg\{\int_{-\beta\mud}^{\infty}\frac{dx~e^{x}}{(e^{x}+1)^2} +\beta\mud\int_{-\beta\mud}^{\infty}\frac{dx~e^{x}}{(e^{x}+1)^2}-\nonumber\\&&~~~-2\beta\mud\int_{-\beta\mud}^{\infty}\frac{dx~e^{2x}}{(e^{x}+1)^3}\bigg\}+\frac{Am\lambda}{6\pi\hbar^2}(\frac{q_{D}}{q_{s}})^3 \frac{q^2_{s}}{\beta}\int_{-\beta\mud}^{\infty}\frac{dx~e^{x}}{(e^{x}+1)^2}+\nonumber\\&& ~~~~~\frac{Aq^4_{s}}{40\pi}(\frac{q_{D}}{q_{s}})^5\bigg\{2\int_{-\beta\mud}^{\infty}\frac{dx~e^{x}}{(e^{x}+1)^3} +2\beta\mud\int_{-\beta\mud}^{\infty}\frac{dx~e^{x}}{(e^{x}+1)^3}\nonumber\\&&~~~~~~~~~~~~~~-4\beta\mud\int_{-\beta\mud}^{\infty}\frac{dx~e^{2x}}{(e^{x}+1)^4}\bigg\}\bigg],
\eeqar
here we have replaced $\sqrt{x+\beta\mud}\simeq \sqrt{\beta\mud}$. Writing the dominating terms containing $\frac{1}{T}$ factor with exponential integrals
\beqar
M^{\p\p}(T\rta 0)&\sim& \frac{1}{T}f_{0}.
\eeqar
 The value of $f_{0}=\int_{-\beta\mud}^{\infty}\frac{dx~e^{x}}{(e^{x}+1)^2}\bigg[1-\frac{2 e^{x}}{(e^{x}+1)}+\frac{2}{(e^{x}+1)}-\frac{4 e^{x}}{(e^{x}+1)} \bigg]$ is constant ($f_{0}=\frac{1}{3})$ in the low temperature limit. Thus we observe power law divergence behavior of quasi localized $d$ or $f$ electrons .  Thus, DC resistivity is given by $\rho(T)=\frac{m}{ne^2} M^{\p\p}(T)=\frac{m}{3ne^2T}$ where $T\ll\frac{\mud}{k_{B}}$.

\subsection{High temperature limit ($k_{B}T\gg\mu_{d}$)}
In high temperature limit  $\beta\mud\ll1$ the general  expression  (\ref{m11}) simplifies to 
\beqar
M^{\p\p}(\om=0,T)&=& \frac{J^2 A^2 m}{12\pi^2N^2nV\hbar^4} \bigg[ \frac{Aq^4_{s}}{40\pi}(\frac{q_{D}}{q_{s}})^5\bigg\{\int_{0}^{\infty}\frac{dx~e^{x}}{(e^{x}+1)^2}+\int_{0}^{\infty}\frac{dx~x~e^{x}}{(e^{x}+1)^2}-2\int_{0}^{\infty}\frac{dx~e^{2x}}{(e^{x}+1)^3}\bigg\}\nonumber\\&&~~+\frac{Am \lambda}{6\pi\hbar^2}(\frac{q_{D}}{q_{s}})^3 \frac{q^2_{s}}{\beta}\int_{0}^{\infty}\frac{dx~e^{x}}{(e^{x}+1)^2}+\frac{Aq^4_{s}}{20\pi}(\frac{q_{D}}{q_{s}})^5\bigg\{\int_{0}^{\infty}\frac{dx~e^{x}}{(e^{x}+1)^3}+\nonumber\\&&~~~~~~~~~~~~~~~~~\int_{0}^{\infty}\frac{dx~x~e^{x}}{(e^{x}+1)^3}-2\int_{0}^{\infty}\frac{dx~x~e^{2x}}{(e^{x}+1)^4}\bigg\}\bigg].
\eeqar
The above expression clearly contains the temperature dependence in the middle term. Thus,
\beqar
M^{\p\p}(k_{b}T\gg\mud)&\simeq& C_{0} T ,
\eeqar
where prefactor $C_{0}=\frac{Am\lambda}{12\pi\hbar^2}(\frac{q_{D}}{q_{s}})^3 q^2_{s}k_{B}$. Therefore, in high temperature limit we observe that memory function varies linearly with temperature. Resistivity is given by $\rho(T)=\frac{m}{ne^2} M^{\p\p}(T)=\frac{ C_{0}mT}{ne^2}$.

\section{2d DC Resitivity in general case }
The temperature dependent DC resistivity can be written from memory function formula (\ref{m11}) using  formula $\rho(T)=\frac{m}{ne^2} M^{\p\p}(T)$ in the general case as:
\beqar
\rho_{2d}(T)&=& (\frac{m}{ne^2})\frac{J^2 A^2 m}{12\pi^2N^2nV\hbar^3} \bigg[\frac{Aq^4_{s}}{40\pi}(\frac{q_{D}}{q_{s}})^5\bigg\{\int_{-\beta\mud}^{\infty}\frac{dx~e^{x}}{(e^{x}+1)^2}+\beta\int_{-\beta\mud}^{\infty}\frac{ dx(\frac{x}{\beta}+\mud)e^x}{(e^{x}+1)^2}\nonumber\\&&~~~-2\beta\int_{-\beta\mud}^{\infty}\frac{ dx(\frac{x}{\beta}+\mud)e^{2x}}{(e^{x}+1)^3}\bigg\}+\frac{Am\lambda}{6\pi\hbar^2}(\frac{q_{D}}{q_{s}})^3\frac{q^2_{s}}{\beta}\int_{-\beta\mud}^{\infty}\frac{dx~e^{x}}{(e^{x}+1)^2}+\nonumber\\&&~~~~~~~\frac{Aq^4_{s}}{20\pi}(\frac{q_{D}}{q_{s}})^5\bigg\{\int_{-\beta\mud}^{\infty}\frac{dx~e^{x}}{(e^{x}+1)^3}+\beta\int_{-\beta\mud}^{\infty}\frac{ dx(\frac{x}{\beta}+\mud)e^x}{(e^{x}+1)^3}\nonumber\\&&~~~~~~~~~~~~~~~~~~~~~-2\beta\int_{-\beta\mud}^{\infty}\frac{ dx(\frac{x}{\beta}+\mud)e^{2x}}{(e^{x}+1)^4}\bigg\}\bigg] .
\eeqar
This is the general formula for  DC resistivity in a $2d$ Kondo lattice system.

\section{Conclusion}
The present work is  the extention of the computation of the DC resistivity of $3d$ Kondo lattice materials. The results are as follows. The temperature of the system is compared  with $\mud$ (chemical potentails of $d$ electrons). At low temperature, I find that DC resistivity proportional to $T^{-1}$. In the high temperature regime resistivity varies linearly with temeperature($\rho\propto T$).  The difference is found only in the high temperature limit(in $3d$ $\rho\propto T^{\frac{3}{2}}$ and $2d$ $ \rho\propto T$).

\appendix
\section*{\bf{Appendix}}

\section{$\phi$ integral solution}
In the presence of Fermi factors of the form $f^s_{k^\p}(1-f^s_{k})$ and at ordinary temperature  $k_{B}T\ll\mus$($\sim$eV), one can replace $\ep$ and $\ep^{\p}$ inside the square root by $\mus$  for $s$ electrons ($\mus=\frac{\hbar^2 q^2_{s}}{2m}$) where $q_{s}$ is Fermi wave vector for $s$-electrons:
\beqar
\mathbb{I}_{\phi}&=&\int_{0}^{2\pi} d\phi \delta (q-\sqrt{2m}\sqrt{(\ep^\p+\ep-2\sqrt{\ep^\p \ep}\cos\phi)})\nonumber\\
&=&\int_{0}^{2\pi}d\phi \delta  \bigg(\underbrace{\frac{\hbar q}{2\sqrt{2m\ep}}}_{h(q,\ep)}- \sin(\frac{\phi}{2})\bigg).\label{phi1}
\eeqar

Using property of delta function $(\delta F(x))=\sum_{} \frac{\delta(x-x_{i})}{|F^\p(x_{i})|}$, $\mathbb{I}_{\phi}$ reduces to

\beqar
\mathbb{I}_{\phi}&=& \frac{\hbar}{2\sqrt{2m\ep}} \int_{0}^{2\pi} d\phi \frac{\delta(\phi-\phi_{0})}{|\frac{1}{2}\cos\frac{\phi}{2}|_{\phi=\phi_{0}}}=\frac{\hbar}{\sqrt{2m\ep}}\frac{1}{\sqrt{1-\frac{q\hbar}{2\sqrt{2m\ep}}}}, \label{phi2}
\eeqar
put $\sqrt{2m\ep}=\hbar q_{s}=\hbar k$ and $\frac{q}{2}=q_{0}$
\beqar
\mathbb{I}({q,k})&=& \frac{1}{q_{s}}\frac{1}{\sqrt{1-\frac{q^2}{4~q^2_{s}}}}=\frac{1}{\sqrt{k^2-q^2_{0}}}.
\label{phi3}
\eeqar
\section{$\mathbb{I}(k^\p)$ integral solution}
\beqar
\mathbb{I}(k^\p)=\int_{0}^{\infty} k^\p dk^\p F(f^s_{k},f^s_{k^\p},f^{d}_{k_{d}},f^d_{k^{\p}_{d}}) [\delta(\ep_{k+q}-\ep_{k}-\hbar\om_{q}+\hbar\om)-\delta(\ep_{k+q}-\ep_{k}-\hbar\om_{q}-\hbar\om)], \nonumber\\ \label{ep1}
\eeqar
changing $k^\p$ integral  into $\ep^\p$ integral by setting $k^\p=\frac{\sqrt{2m\ep^\p}}{\hbar}$ and $dk^\p=\frac{\sqrt{2m}}{2\hbar}\frac{d\ep^\p}{\sqrt{\ep^\p}}$
\beqar
\mathbb{I}(\ep^\p)&=& \frac{m}{\hbar^2}\int_{0}^{\infty} d\ep^\p\bigg\{ f^{s}(\ep_{k+q}) (1-f^{s}(\ep_{k}))\sum_{k_{d}}F^{1}_{d}(q)-(f^{s}(\ep_{k})-f^{s}(\ep_{k+q}))\sum_{k_{d}}F^{2}_{d}(q)\bigg\}\nonumber\\&&~~~~~~~~~~~~~~~~~[\delta(\ep_{k+q}-\ep_{k}-\hbar\om_{q}+\hbar\om)-\delta(\ep_{k+q}-\ep_{k}-\hbar\om_{q}-\hbar\om)] ,\label{ep2}
\eeqar
here $F^{1}_{d}(q)=\sum_{k_{d}}(f^{d}_{k_{d}}-f^{d}_{k^{\p}_{d}})$, $F^{2}_{d}(q)=\sum_{k_{d}} f^{d}_{k_{d}}(1-f^{d}_{k^{\p}_{d}})$ and $\ep^\p=\ep_{k+q}$. The property of delta function $\int dx f(x)\delta(x-a)= f(a)$
changes the above expression to
\beqar
\mathbb{I}(\ep)&=&\frac{m}{\hbar^2}\bigg[\bigg\{f^{s} (\ep_{k}+\hbar\om_{q}-\hbar\om) \bigg(1-f^s(\ep_{k})\bigg)F^{1}_{d}(q)-\bigg(f^{s} (\ep_{k})-f^s(\ep_{k}+\hbar_{q}-\hbar\om)\bigg) F^{2}_{d}(q) \bigg\} \nonumber\\&&-\bigg\{f^{s} (\ep_{k}+\hbar\om_{q}+\hbar\om) \bigg(1-f^s(\ep_{k})\bigg)F^{1}_{d}(q) -\bigg(f^{s} (\ep_{k})-f^s(\ep_{k}+\hbar \om_{q}+\hbar\om)\bigg) F^{2}_{d}(q)\bigg\}  \bigg], \nonumber\\&& \label{ep3}
\eeqar
simplifies to
\beqar
\mathbb{I}(\ep)&=& \frac{m}{\hbar^2}\bigg[\bigg\{f^{s} (\ep_{k}+\hbar\om_{q}-\hbar\om)-f^{s} (\ep_{k}+\hbar\om_{q}+\hbar\om)\bigg\} \bigg(1-f^s(\ep_{k})\bigg)F^{1}_{d}(q)+\nonumber\\&&~~~~~~~~~~\bigg\{f^{s} (\ep_{k}+\hbar\om_{q}-\hbar\om)-f^{s} (\ep_{k}+\hbar\om_{q}+\hbar\om) \bigg\}F^{2}_{d}(q)\bigg].
\eeqar

\section{Computation of  $\mathbb{I}_{1}(q)$  }

\beqar
\mathbb{I}_{1}(q)=
\int_{0}^{q_{D}} dq q^2\sum_{k_{d},k^{\p}_{d}}(f^{d}_{k_{d}}-f^{d}_{k^{\p}_{d}}) \label{iq0}
\eeqar
write
\beqar
f^1_{d}(q)&=&\sum_{k_{d}}[ f^{d}(\ep_{k_{d}})-f^d(\ep_{k^{\p}_{d}})]. \label{fd0}
\eeqar 

Small $q$ expansion of $f^1_{d}(q)$ gives
\beqar
f^1_{d}(q)&=&-\frac{A}{(2\pi)^2}\int_{0}^{\infty}k_{d} dk_{d}\int_{0}^{2\pi} d\phi\bigg[q\frac{\pr f^{d}(\ep_{k^{\p}_{d}})}{\pr q}|_{q=0}+\frac{q^2}{2!}\frac{\pr^2 f^{d}(\ep_{k^{\p}_{d}})}{\pr q^2}|_{q=0}\nonumber\\&&~~~~~~~~~~~~~~~~~~~~~~~+\frac{q^3}{3!}\frac{\pr^3 f^{d}(\ep_{k^{\p}_{d}})}{\pr q^3}|_{q=0}...\bigg].\nonumber\\  \label{fd2}
\eeqar
We have Fermi function $f^{d}(\ep_{k^{\p}_{d}},\theta)=\frac{1}{e^{\beta[\frac{\hbar^2 q^2}{2m_{d}}+\frac{\hbar^2 k_{d}^2}{2m_{d}}+\frac{\hbar^2 k_{d} q \cos\theta}{m_{d}]}-\mu_{d}]}+1}$. For simplification, On performing Taylor's expansion for small ($ q \rta 0 $) and converting summation into integral we getwe put $\alpha=\beta(\frac{\hbar^2 k_{d}^2}{2m_{d}}-\mu_{d})$, $\eta=\beta\frac{\hbar^2}{2m_{d}}$ and $\gamma=\beta\frac{\hbar^2 k_{d} }{m_{d}}$. The Fermi function set to
\beqar 
f^d(q,\alpha,\eta,\gamma,\theta)=\frac{1}{e^{[\alpha+\eta q^2+\gamma q \cos\theta]}}~,~~~~\frac{\pr f^d(\alpha,\gamma,\theta)}{\pr q}|_{q=0}=-\frac{e^{\alpha}\gamma \cos\theta }{(e^{\alpha}+1)^2},\nonumber\\  \label{fd3}
\eeqar
on substituting derivatives of fermi functions for small q limit and performing $\phi$ integral eqn (\ref{fd2}) becomes
\beqar
f^1_{d}(q,k_{d})&=&\frac{q^2A}{2 (2\pi)^2}\int_{0}^{\infty}k_{d} dk_{d}\bigg\{ \frac{e^{\alpha}}{(e^{\alpha}+1)^2}[4\pi\eta+\gamma^2\pi]-\frac{2\pi \gamma^2e^{2\alpha}}{(e^{\alpha}+1)^3}+..... \bigg\}. \label{fd4}
\eeqar
We convert $k_{d}$ integral into energy integral ($\ep_{d}$) and replace $\alpha$,$\beta$ and $\gamma$ with their respective terms
\beqar
f^1_{d}(q,\ep_{d})&= &\frac{A m q^2}{8\pi\hbar^2} \bigg[\frac{2\beta\hbar^2}{m}\int_{0}^{\infty}\frac{d\ep_{d}e^{\beta(\ep_{d}-\mud)}}{(e^{\beta(\ep_{d}-\mud)}+1)^2}+\frac{2\beta^2\hbar^2}{m}\int_{0}^{\infty}\frac{d\ep_{d}\ep_{d}e^{\beta(\ep_{d}-\mud)}}{(e^{\beta(\ep_{d}-\mud)}+1)^2}\nonumber\\&&~~~~~~~~~~~~~~~~~~~~~~~~~~-\frac{4\beta^2\hbar^2}{m}\int_{0}^{\infty}\frac{d\ep_{d}\ep_{d}e^{\beta(\ep_{d}-\mud)}}{(e^{2\beta(\ep_{d}-\mud)}+1)^3} \bigg],\nonumber\\ \label{fd5}
\eeqar
simplifies to
\beqar
f^1_{d}(q,\ep_{d})&=& \frac{Aq^2}{4\pi}\bigg[\beta\int_{0}^{\infty}\frac{d\ep_{d}e^{\beta(\ep_{d}-\mud)}}{(e^{\beta(\ep_{d}-\mud)}+1)^2}+\beta^2\int_{0}^{\infty}\frac{d\ep_{d}\ep_{d}e^{\beta(\ep_{d}-\mud)}}{(e^{\beta(\ep_{d}-\mud)}+1)^2}-2\beta^2\int_{0}^{\infty}\frac{d\ep_{d}\ep_{d}e^{2\beta(\ep_{d}-\mud)}}{(e^{\beta(\ep_{d}-\mud)}+1)^3}\bigg].~~~\nonumber\\&&~ \label{fd6}
\eeqar
Substituing the above expression in eqn (\ref{iq0}) and performing integration over $q$ we obtain
\beqar
\mathbb{I}_{1}(q)&=&
\frac{Aq^5_{D}}{20\pi}\bigg[\beta\int_{0}^{\infty}\frac{d\ep_{d}e^{\beta(\ep_{d}-\mud)}}{(e^{\beta(\ep_{d}-\mud)}+1)^2}+\beta^2\int_{0}^{\infty}\frac{d\ep_{d}\ep_{d}e^{\beta(\ep_{d}-\mud)}}{(e^{\beta(\ep_{d}-\mud)}+1)^2}-2\beta^2\int_{0}^{\infty}\frac{d\ep_{d}\ep_{d}e^{2\beta(\ep_{d}-\mud)}}{(e^{\beta(\ep_{d}-\mud)}+1)^3}\bigg]. \nonumber\\ \label{iq2}
\eeqar
\section{Computation of  $\mathbb{I}_{2}(q)$  }
\beqar
\mathbb{I}_{2}(q)=\int_{0}^{q_{D}} dq q^2\sum_{k_{d},k^{\p}_{d}}f^{d}_{k_{d}}(1-f^{d}_{k^{\p}_{d}}), \label{iq1}
\eeqar
write
\beqar
f^2_{d}(q)&=&\sum_{k_{d}} f^{d}(\ep_{k_{d}})[1-f^d(\ep_{k^{\p}_{d}})]. \label{fd01}
\eeqar 
On performing Taylor's expansion for small ($ q \rta 0 $) and converting summation into integral we get
\beqar
f^2_{d}(q)&=& \frac{A}{(2\pi)^2}\int_{0}^{\infty}k_{d} dk_{d}\int_{0}^{2\pi}d\phi  f^{d}(\ep_{k_{d}})(1- f^{d}(\ep_{k_{d}}))-\frac{A\pi}{(2\pi)^2}\frac{q^2}{2}\int_{0}^{\infty}k_{d} dk_{d}\bigg\{\frac{-e^{\alpha}}{(e^{\alpha}+1)^3}\times\nonumber\\&&~~~~~~~~~~~~~~~~~~~~~~~~~~~~~~~~~~~~~~~~~ [4\eta+\gamma^2 ] +2\gamma^2\frac{e^{2\alpha}}{(e^{\alpha}+1)^4}\bigg\}. \label{fd02}
\eeqar
We convert $k_{d}$ integral into energy integral ($\ep_{d}$) and replace $\alpha$,$\beta$ and $\gamma$ with their respective terms
\beqar
f^2_{d}(q,\ep_{d})&=& \frac{Am_{d}}{2\pi\hbar^2}\int_{0}^{\infty}\frac{d\ep_{d}e^{\beta(\ep_{d}-\mud)}}{(e^{\beta(\ep_{d}-\mud)}+1)^2}+\frac{Aq^2}{4\pi}\bigg[\beta\int_{0}^{\infty}\frac{d\ep_{d}e^{\beta(\ep_{d}-\mud)}}{(e^{\beta(\ep_{d}-\mud)}+1)^3}+\beta^2\int_{0}^{\infty}\frac{d\ep_{d}\ep_{d}e^{\beta(\ep_{d}-\mud)}}{(e^{\beta(\ep_{d}-\mud)}+1)^3}\nonumber\\&&~~~~~~~~~~~~~~-2\beta^2\int_{0}^{\infty}\frac{d\ep_{d}\ep_{d}e^{2\beta(\ep_{d}-\mud)}}{(e^{\beta(\ep_{d}-\mud)}+1)^4}\bigg]. \label{fd03}
\eeqar
On substituting the above expression into eqn(\ref{iq1}) and performing $q$ integration we get 
\beqar
\mathbb{I}_{2}(q)&=&\frac{Am_{d}q^3_{D}}{6\pi\hbar^2}\int_{0}^{\infty}\frac{d\ep_{d}e^{\beta(\ep_{d}-\mud)}}{(e^{\beta(\ep_{d}-\mud)}+1)^2}+ \frac{Aq^5_{D}}{20\pi}\bigg[\beta\int_{0}^{\infty}\frac{d\ep_{d}e^{\beta(\ep_{d}-\mud)}}{(e^{\beta(\ep_{d}-\mud)}+1)^3}+\beta^2\int_{0}^{\infty}\frac{d\ep_{d}\ep_{d}e^{\beta(\ep_{d}-\mud)}}{(e^{\beta(\ep_{d}-\mud)}+1)^3}\nonumber\\&&~~~~~~~~~~~~~~-2\beta^2\int_{0}^{\infty}\frac{d\ep_{d}\ep_{d}e^{2\beta(\ep_{d}-\mud)}}{(e^{\beta(\ep_{d}-\mud)}+1)^4}\bigg]. \nonumber\\&&  \label{iq22}
\eeqar

\section*{Acknowledgement}
I sincerely thank Dr. Navinder Singh and Prof. Raman Sharma for providing their invaluable guidance and  stimulating discussions.

\end{document}